\documentclass[aoas,preprint]{imsart}

\usepackage{bm}
\RequirePackage[OT1]{fontenc}
\RequirePackage{amsthm,amsmath}
\RequirePackage{natbib}
\setcitestyle{authoryear,open={(},close={)}}

\usepackage{graphicx}
 \usepackage{multirow}
 \usepackage{lscape}
\usepackage{tablefootnote}
\usepackage[normalem]{ulem}
\usepackage{amssymb}
\usepackage[ruled,vlined,linesnumbered]{algorithm2e}
\usepackage{xcolor}
\usepackage{soul}
\usepackage{amsmath}
\usepackage{enumerate}
\usepackage{natbib}
\usepackage{url} 
\usepackage{ulem}
\newcommand{\stkout}[1]{\ifmmode\text{\sout{\ensuremath{#1}}}\else\sout{#1}\fi}

\usepackage{lineno}
\arxiv{arXiv:0000.0000}

\startlocaldefs
\numberwithin{equation}{section}
\theoremstyle{plain}

\endlocaldefs

\begin{document}

\begin{frontmatter}
\title{Bayesian Non-Homogeneous Hidden Markov Model with Variable Selection for Investigating Drivers of Seizure Risk Cycling}
\runtitle{Hidden Markov Model for Seizure Risk Assessment}

\begin{aug}
\author{\fnms{Emily T.} \snm{Wang}\thanksref{t1}\ead[label=e1]
{emilywang@rice.edu}},

\author{\fnms{Sharon} \snm{Chiang}\thanksref{t2}\ead[label=e2]
{sharon.chiang@ucsf.edu}},

\author{\fnms{Zulfi} \snm{Haneef}\thanksref{t3}\ead[label=e3]
{zulfi.haneef@bcm.edu}},

\author{\fnms{Vikram R.} \snm{Rao}\thanksref{t2}\ead[label=e4]
{vikram.rao@ucsf.edu}},
 
\author{\fnms{Robert} \snm{Moss}\thanksref{t4}\ead[label=e5]
{rob@seizuretracker.com}},
       
\and\\
\author{\fnms{Marina} \snm{Vannucci}\thanksref{t1}\ead[label=e6]{marina@rice.edu}}

\runauthor{E. T. Wang et al.}

\affiliation{Rice University\thanksmark{t1} and  University of California, San Francisco\thanksmark{t2} and Baylor College of Medicine\thanksmark{t3} and SeizureTracker LLC\thanksmark{t4}}

\address{Department of Statistics\\
Rice University\\
Houston, TX, USA\\
\printead{e1}\\ \printead{e6} }

\address{Department of Neurology\\ 
University of California, San Francisco\\
San Francisco, CA, USA\\
\printead{e2}\\ \printead{e4} }

\address{Department of Neurology\\ 
Baylor College of Medicine\\
Houston, TX, USA\\
\printead{e3} }

\address{Seizure Tracker, LLC\\ 
Springfield, VA, USA\\
\printead{e5} }

\end{aug}

\begin{abstract}
A major issue in the clinical management of epilepsy is the unpredictability of seizures. Yet, traditional approaches to seizure forecasting and risk assessment in epilepsy rely heavily on raw seizure frequencies, which are a stochastic measurement of seizure risk. We consider a Bayesian non-homogeneous hidden Markov model for unsupervised clustering of zero-inflated seizure count data. The proposed model allows for a probabilistic estimate of the sequence of seizure risk states at the individual level. It also offers significant improvement over prior approaches by incorporating a variable selection prior for the identification of clinical covariates that drive seizure risk changes and accommodating highly granular data. For inference, we implement an efficient sampler that employs stochastic search and data augmentation techniques. We evaluate model performance on simulated seizure count data. We then demonstrate the clinical utility of the proposed model by analyzing daily seizure count data from 133 patients with Dravet syndrome  collected through the {\it Seizure Tracker$^{TM}$} system, a patient-reported electronic seizure diary. We report on the dynamics of seizure risk cycling, including validation of several known pharmacologic relationships. We also uncover novel findings characterizing the presence and volatility of risk states in Dravet syndrome, which may directly inform counseling to reduce the unpredictability of seizures for patients with this devastating cause of epilepsy.
\end{abstract}


\end{frontmatter}

\section{Introduction}
Epilepsy, a chronic neurological disorder, is characterized by frequent, unpredictable seizures arising from abnormal electrical disturbances in the brain. This disorder affects over 60 million people worldwide, which is approximately equivalent to a 1\% disease prevalence \citep{who2019}. The unpredictable nature of seizures not only makes treatment of epilepsy difficult but leads to increased morbidity and mortality and severely reduces patients’ quality of life \citep{arthurs2010patient}.  Currently, clinical decision-making in epilepsy depends heavily on raw seizure counts and decisions about treatment are based primarily on whether the seizure frequency has increased or decreased after an intervention. It is recognized, however, factors other than treatment, such as cycles in underlying epileptiform discharges, may modulate the likelihood of seizures \citep{baud2020}. 

It is increasingly recognized that seizures are stochastic realizations of periods of heightened seizure risk \citep{chiang2018,baud2020}; given any underlying seizure risk level, the overt number of seizures that a patient has can vary due to natural probabilistic variation and the inherent unpredictability of seizures, making clinical prediction of seizures challenging \citep{goldenholz2018, chiang2018}. Consequently, raw seizure counts are only a surrogate measure and recognized not to be an accurate measure of a patient's true seizure risk. This concept was first formalized by \cite{chiang2018}, who introduced the notion of discrete unknown seizure risk ``states'' and showed that seizure risk can be estimated as a latent quantity based on the seizure counts. 
The validity of this conceptual approach to seizure risk has been validated against specialized epilepsy clinician experts \citep{chiang2020} and confirmed by empiric observations from chronic intracranial electrocorticography data, which have demonstrated that seizures tend to occur at specific phases of underlying fluctuations in interictal epileptiform activity \citep{baud2018,karoly2018,rao2020cues,proix2020forecasting,leguia2021seizurecycles}.

Here, we build upon the literature described above by proposing a Bayesian non-homogeneous hidden Markov model for zero-inflated count data that yields substantially increased flexibility for accommodating highly granular data and that allows for simultaneous selection of high-dimensional covariates to identify drivers of seizure risk cycles. To handle high proportions of zeros as well as overdispersion, we model daily seizure counts using a zero-inflated negative binomial (ZINB) distribution. We incorporate the external clinical covariates into the estimation of both emission and transition probabilities through logistic regression frameworks that allow additional flexibility across different subjects and time points, and employ variable selection priors to simultaneously identify the significant covariates. The proposed model offers significant advances as it allows for (1) high temporal granularity of the count data,  (2) fine-tuned and personalized risk assessment via the modeling of the effect of subject-level clinical factors on both state transitions and the current expected number of seizures, and (3) variable selection to simultaneously identify covariates that drive seizure risk changes. These aspects are essential to developing a generalizable model capable of modeling seizure risk data in diverse datasets. 

We optimize inference by implementing a Markov chain Monte Carlo method that uses stochastic search methods and data augmentation techniques for efficient sampling \citep{savitsky2011,polson2013}. We evaluate model performance on simulated seizure count data. We then demonstrate the clinical utility of our model by analysing daily clinical seizure counts recorded by patients with a severe genetic cause of epilepsy, Dravet syndrome (DS), a catastrophic developmental epileptic encephalopathy affecting one in 15,700 infants \citep{wu2015incidence}. Infants with DS often present with the first convulsive seizure within the first year of life. Shortly thereafter, multiple other seizure types emerge; eventually, patients develop refractory convulsive seizures, multiple seizure types, and intellectual disability. Mortality in Dravet syndrome is high and up to 17\% of patients with Dravet syndrome die before adulthood, with 15--61\% of deaths attributable to sudden unexpected death in epilepsy (SUDEP) \citep{cooper2016}.
The single greatest risk factor for SUDEP is the presence of generalized tonic-clonic seizures (GTCs) \citep{harden2017practice}. Patients with more than three GTCs in the preceding year have more than an eight-fold increased risk of death from SUDEP \citep{walczak2001incidence}. Understanding patterns in GTC cycling as well as which anti-seizure medications (ASMs) and triggers reduce or increase the likelihood of transitioning to periods of lowered seizure risk for GTCs is of interest to guide understanding and prevention of SUDEP \citep{ayub2020natural}. However, the stochastic nature of seizures makes this task difficult.

In our analysis, we consider data from 133 patients with Dravet syndrome from one of the world's largest seizure diary databases, Seizure Tracker LLC\footnote{
Seizure Tracker™ - Your comprehensive resource for tracking and sharing seizure information, \url{https://seizuretracker.com/}.}, and a total of 36 potential variables that may drive seizure risk. Our approach identifies the presence of three distinct seizure risk states in Dravet syndrome through which patients cycle, that roughly equate to a states reflecting low, moderate, and high propensity for GTCs. We report on the volatility and time dynamics of these states in order to contribute to knowledge on the natural history of seizure risk cycling in Dravet syndrome. Lastly, we show that our model accurately recovers several known pharmacologic relationships of ASMs in Dravet syndrome. This application improves understanding of seizure risk cycling in Dravet syndrome and illustrates the usefulness of our approach more broadly for individualized  investigations of seizure risk cycling in epilepsy. 

In Section 2 below we detail the proposed model and the inference strategy.  In Section 3 we utilize simulated seizure count data to evaluate performances of our model. In Section 4 we report on the results from the analysis of daily seizure counts data from patients with Dravet syndrome and discuss our findings. We give concluding remarks in Section 5.

\section{Methods}
\begin{figure}
\centering
\caption{{\bf Proposed hidden Markov model for seizure risk estimation (ZINB-NHMM-BVS).}
Depending on exogenous covariates, patients transition between different seizure risk states from day to day. Then, conditional upon the states, seizure counts are emitted from zero-inflated negative binomial distributions, which are also dependent on covariates.}
\includegraphics[scale = 0.65]{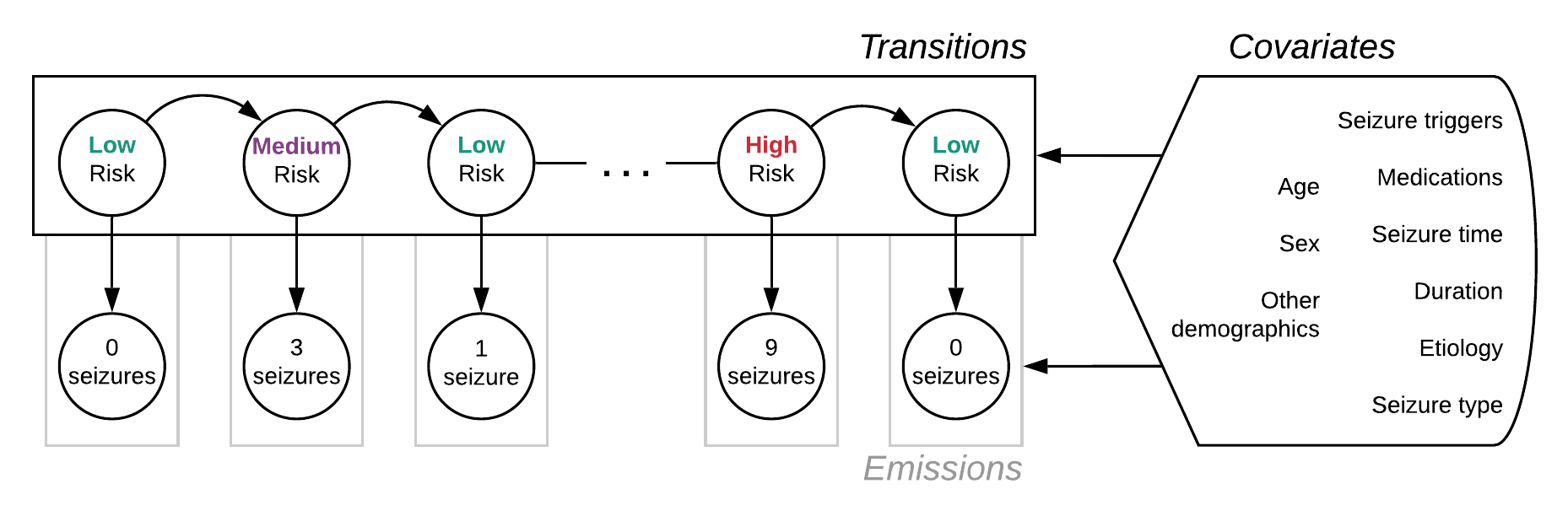}
\label{clinicalgraph}
\end{figure}

A hidden Markov Model (HMM) is a statistical model which assumes that a system is in one of many latent (unobservable) states at any point in time \citep{rabiner1989}. The system transitions from one state to another over time, and observations are generated from an emission distribution, conditional on the latent state sequence. Fig~\ref{clinicalgraph} depicts our proposed HMM, which we call ZINB-NHMM-BVS, for seizure risk assessment. The latent layer allows a probabilistic estimate of the sequence of seizure risk states at an individual level. The process in which observed seizure counts are emitted from a statistical distribution dependent on the risk state also allows for a degree of natural variance in counts to be accounted for.  Our model formulation additionally incorporates clinical covariates and employs variable selection to perform simultaneous inference on variables significantly associated with changes in latent risk. 

We now describe our model, choice of priors and MCMC algorithm for posterior inference in detail.

\subsection{Bayesian Non-Homogeneous HMM for Zero-Inflated Count Data}

We assume to have data observed on $N$ patients, with each patient $i \in \{1, \ldots, N\}$ recording daily seizure counts over $T_i$ days. Let $Y_{it}\in\{0, 1, \ldots\}$ be the number of seizures of patient $i$ at time $t$, for $i=1, \ldots,N$ and $t=1, \ldots,T_{i}$. The number of days recorded per patient and the initial starting day for each patient do not need to be the same. Let $\mathbf{W}_{it}=(W_{it1}, \ldots,W_{itp})$ be the values of $p$ time-varying covariates for patient $i$ at time $t$, and $\mathbf{Z}_{i}=(Z_{i1}, \ldots,Z_{iq})$ the values of $q$ fixed covariates. Finally, let $\mathbf{X}_{i}=(\mathbf{W}_{i},\mathbf{1}_{q\times1}\mathbf{Z}_{i}$) be the $((q+p)\times T_{i})$ stacked matrix of all clinical covariates for patient $i$.

To accommodate the temporal structure in the seizure count process, we model the seizure counts as a $K$-state non-homogeneous first-order HMM, where $K \in \mathbb{N}$. Let $\xi_{it}$ denote the latent seizure risk state of patient $i$ at time $t$. By the Markov property, the probability of patient $i$ transitioning to a specific state $k \in \{1, \ldots, K\}$ at time $t$ depends only upon the hidden state at the previous time point, $\xi_{i(t-1)}$. Furthermore, in a non-homogeneous HMM, the transition probabilities vary over time as a function of the covariates. Specifically, we use a multinomial logit regression to represent these transition probabilities as
\begin{equation}
\label{multilogit}
\mathrm{Pr}(\xi_{it}=k\mid\xi_{i,t-1}=k',\boldsymbol{X}_{i,t-1},\boldsymbol{\beta}_{k'})  =\frac{\exp(\boldsymbol{X}_{i,t-1}^{T}\boldsymbol{\beta}_{k'k})}{1+\sum_{l=1}^{K-1}\exp(\boldsymbol{X}_{i,t-1}^{T}\boldsymbol{\beta}_{k'l})}. 
\end{equation}
The use of logistic regression to link covariate effects to transition probabilities was first proposed by \cite{muenz1985}, for the case of a two-state Markov chain. To maintain model identifiability, we establish a baseline state and set to zero the regression coefficients corresponding to transitions into this state. With this representation, the coefficient $\boldsymbol{\beta}_{j,k'k}$ measures the effect of covariate $j$ on the {\it relative} likelihood of patients transitioning from state $k'$ to state $k$, compared to transitioning from state $k'$ to the predetermined baseline state. Positive (negative) values of $\boldsymbol{\beta}_{j,k'k}$ indicate that covariate $j$ increases (decreases) the chance of patients transitioning from state $k'$ to $k$ over transitioning from state $k'$ to the baseline. Each individual transition from one state to another has its own set of regression coefficients. Thus, for $p$ covariates there are a total of $K \times (K-1) \times p$ coefficients governing the Markov transitions. An intercept may be modeled by appending a vector of ones to the data matrix $\boldsymbol{X}$. We notice that using fixed regression coefficients for transitions from any state into a new state $k'$ in Eq~\eqref{multilogit} would lead to a more parsimonious model, see for example \cite{holsclaw2017}. However, in our case study, separate sets of coefficients for each transition are essential as we are primarily interested in identifying covariates associated with escalations or de-escalations of seizure risk. In Section ``Variable Selection Priors'' below we employ specialized priors to achieve this objective.

Next we describe the emission distribution of our HMM model. Count data is typically modeled using the Poisson, binomial, or negative binomial distributions. The Poisson distribution is constrained in that the expected value must be equal to the variance. However, daily seizure counts usually exhibit extreme zero-inflation and overdispersion in which the variance exceeds the mean \citep{tharayil2017}. Therefore, we assume that seizure counts at each state $k$ follow a zero-inflated negative binomial (ZINB) mixture distribution which assumes that, conditional upon being in state $k$, the observed count is either zero with probability $p_k$, or it comes from a negative binomial random variable, with probability $1-p_k$. To facilitate computation, we introduce auxiliary variables $Z_{it}$ that allow us to keep track of whether an observation belongs to the negative binomial mixture component or the ``zero" component, that is $Z_{it}\mid p_{k}\sim \mathrm{Bernoulli}(p_{k})$ where $Z_{it}=1$ if $Y_{it}=0$ and $Z_{it}=0$ if $Y_{it}$ is negative binomial. We follow \cite{pillowscott2012} and parametrize the negative binomial distribution in terms of state-specific dispersion parameters $r_k$ and subject- and state-dependent probabilities $\psi_{itk}$ (equivalent to $1 - $ success probability) as
\begin{equation}
\label{negbin}
p(Y_{it}\mid\psi_{itk},\boldsymbol{r},\xi_{it}=k,Z_{it}=0) =\frac{\Gamma(Y_{it}+r_{k})}{\Gamma(r_{k})Y_{it}!}(1-\psi_{itk})^{r_{k}}\psi_{itk}^{Y_{it}}.
\end{equation}
This parametrization, in particular, allows us to incorporate both subject-level covariates and time-varying effects into the emission distributions by letting the probabilities $\psi_{itk}$ depend on the covariates as 
\begin{equation}
\label{psilogit}
\mathrm{\psi}_{itk} =\frac{exp(\boldsymbol{X}_{it}^{T}\boldsymbol{\rho}_{k})}{1+exp(\boldsymbol{X}_{it}^{T}\boldsymbol{\rho}_{k})},
\end{equation}
with $\rho_{k}$ a state-dependent vector of regression coefficients. A separate set of covariates than those used for the transition probability regression can be considered, if desired. Parametrizing the negative binomial distribution in terms of $\psi_{itk}$ instead of the mean yields a closed-form full conditional posterior for the regression coefficients $\boldsymbol{\rho}$. Mean parameters can be easily recovered as
$\mu_{itk}= \frac{\psi_{itk}\,r_{k}}{1-\psi_{itk}}$. We can now write our ZINB model as
\begin{equation}
\label{meanparam}
[Y_{it}\mid\xi_{it}=k, \mathbf{r},\psi_{itk},\mathbf{p}]\:\sim\: p_{k}\,1_{\{Y_{it}=0\}}+(1-p_{k})\,\mathrm{NB}(r_{k},\psi_{itk}),
\end{equation}
with $\mathrm{NB}(r_{k},\psi_{itk})$ given by Eq~\eqref{negbin}. 

\subsection{Variable Selection Priors}
\label{Variable Selection Priors}

We are interested in identifying clinical covariates which are associated with changes in seizure risk. We achieve this task by employing variable selection priors \citep{george1997,brown1998}. Let us first consider the transition probabilities \eqref{multilogit}.  We introduce a $K \times (K-1) \times p$ latent inclusion tensor $\gamma$ such that $\gamma_{j,k'k} = 1$ if variable $j$ has a non-zero effect on the likelihood of transitioning from state $k'$ to state $k$. We then impose a {\it spike-and-slab} prior on the regression coefficient $\beta_{j,k'k}$, obtained as a mixture of a Gaussian distribution (the slab) and a point mass at zero (the spike) as
\begin{equation}
\label{betaspikeandslab}
\left[\beta_{j,k'k}\mid\gamma_{j,k'k},\mu_{\beta},\sigma_{\beta}^2\right]\sim\gamma_{j,k'k} N(\mu_{\beta},\sigma_{\beta}^{2})+(1-\gamma_{j,k'k})\delta_{0}(\beta_{j,k'k}).
\end{equation}
We further assume $\gamma_{j,k'k}\sim$ Bernoulli$(\theta_{j,k'k})$, with a hyperprior on the inclusion probability $\theta_{j,k'k} \sim \mathrm{Beta}(g_{\beta},h_{\beta})$, and then integrate $\theta_{j,k'k}$ out. Multivariate {\it spike-and-slab} prior constructions of type \eqref{betaspikeandslab} have been employed in multivariate regression models to identify covariates that are associated with individual responses \citep{stingo2010}. In our framework, this construction allows us to select covariates that are associated with specific transitions, from one risk state to another one. We note that these inferences are conducted in relation to the baseline state. Selection of covariate $j$ as important for transitions from state $k'$ to state $k$ implies that the value of covariate $j$ has a significant effect on the likelihood of a patient transitioning from state $k'$ to state $k$, over state $k'$ to the baseline state.

We adopt a similar prior setup for the regression coefficients $\boldsymbol{\rho}_k$ of the state-dependent probabilities \eqref{psilogit}. Let $\delta_{jk}$ be a latent indicator of whether variable $j$ has a non-zero effect on the probability parameter of state $k$, $\psi_k$. We impose a {\it spike-and-slab} prior on $\rho_{jk}$ of the type
\begin{equation}
\label{rhospikeandslab}
\left[\rho_{jk}\mid\delta_{jk},\mu_{\rho},\sigma_{\rho}^2\right] \sim\delta_{jk} N(\mu_{\rho},\sigma_{\rho}^{2})+(1-\delta_{jk})\delta_{0}(\rho_{jk}),
\end{equation}
with $\delta_{jk}\sim$ Bernoulli$(\theta_{jk})$ and a hyperprior $\theta_{jk} \sim \mathrm{Beta}(g_{\rho},h_{\rho})$. The two variable selection schemes outlined above offer separate interpretations. Variable selection for the emission distributions via $\rho$ determines which covariates are associated with higher or lower seizure frequency, conditional on the underlying latent state. In contrast, variable selection for the transitions identifies covariates which are associated with worsening or improvement of seizure risk, from time $t$ to $t+1$.

Lastly, we specify a Beta prior for the zero-inflation parameters, $p_{k}\sim \mathrm{Beta}(c,d)$ for $k=1,\ldots,K$, a Gamma prior on the dispersion parameters, $r_{k}\sim \mathrm{Gamma}(e,f)$, and assume the distribution of the initial hidden state at time $t=1$ to be $\xi_{i1}\sim \mathrm{Multinomial}(1;\pi_{1},...,\pi_{K})$, with
a conjugate Dirichlet prior on $\boldsymbol{\pi}$.

\subsection{Data Augmentation}
\label{sec:dataaug}

Given the model construction, the full likelihood is 
\begin{align*}
\mathfrak{L}(\mathbf{Y}\mid\boldsymbol{\xi} & ,\boldsymbol{r},\boldsymbol{\psi},\boldsymbol{p},\boldsymbol{X},\boldsymbol{\beta})\propto\prod_{i=1}^{N}p(\xi_{i1})\,\times\prod_{i=1}^{N}\prod_{t=1}^{T_{i}}\frac{exp(\boldsymbol{X}_{i,t-1}^{T}\boldsymbol{\beta}_{\xi_{i,t-1},\xi_{it}})}{1+\sum_{l=1}^{K-1}exp(\boldsymbol{X}_{i,t-1}^{T}\boldsymbol{\beta}_{\xi_{i,t-1},l})}\\
& \times\,\prod_{i=1}^{N}\prod_{t=1}^{T_{i}}\left[(1-p_{\xi_{it}})+p_{\xi_{it}}(1-\psi_{it\xi_{it}})^{r_{\xi_{it}}}\right]^{1-Z_{it}}\\
& \times\left[p_{\xi_{it}}\frac{\Gamma(Y_{it}+r_{\xi_{it}})}{\Gamma(r_{\xi_{it}})Y_{it}!}(1-\psi_{it\xi_{it}})^{r_{\xi_{it}}}\psi_{it\xi_{it}}^{Y_{it}}\right]^{Z_{it}}.
\end{align*}
In our sampling scheme for posterior inference, we make use of modern data augmentation techniques that allow us to optimize the inference.

\subsubsection{Inference on the transition matrix elements}
\label{section2.3.1}

In our model, the probability of a patient transitioning from risk state $k$ on day $t$ to state $k'$ on the next day, $t+1$, is dependent on the values of the patient's clinical covariates on the original day $t$ via the multinomial logistic regression of Eq~\eqref{multilogit}. Sampling schemes for the multinomial logit regression coefficients are challenging to implement, due to the intractable form of the likelihood and the lack of a conjugate prior for the coefficients. An efficient data augmentation approach for multinomial logistic regression models via P\'{o}lya-Gamma distributions was proposed by \cite{polson2013}. This method essentially manipulates terms in the full likelihood such that they combine into a single Gaussian kernel in the posterior. The key result which makes P\'olya-Gamma random variables useful in the logistic setting is that for $b>0$,
\begin{align}
\label{pgkey}
\frac{(e^{\psi})^{a}}{(1+e^{\psi})^{b}} & =2^{-b}e^{\kappa\psi}\int_{0}^{\infty}e^{-\omega\psi^{2}/2}p(\omega)\,d\omega,
\end{align}
where $\kappa=a-b/2$ and $\omega\sim\mathrm{PG}(b,0)$.  

Let $Y_{itk}= I_{\xi_{it}}(k)$ be a binary representation of the latent states, where $I_{\xi_{it}}(k)$ is an indicator function equal to 1 if $\xi_{it}=k$ and 0 otherwise. Then using the multinomial logistic representation of transition probabilities in Eq~\eqref{multilogit} and following \cite{holmes2006}, the posterior can be rewritten as 
\begin{align*}
p(\boldsymbol{\beta}\mid\boldsymbol{X},\boldsymbol{Y}) & =p(\boldsymbol{\beta})\cdot\prod_{i=1}^{N}\prod_{t=2}^{T_{i}}\prod_{k=1}^{K}\left[\frac{exp(\zeta_{itk}-C_{itk})}{1+exp(\zeta_{itk}-C_{itk})}\right]^{Y_{itk}},
\end{align*}
where $\zeta_{itk}=\boldsymbol{X}_{i,t-1}^{T}\boldsymbol{\beta}_{k'k}$ and $C_{itk}=log\sum_{j\neq k}exp\left\{ \zeta_{itj}\right\}$. Making use of the property of the P\'olya-Gamma distribution in Eq~\eqref{pgkey}, the full conditional posterior of $\boldsymbol{\beta}_{k'k}$, given all other parameters of the model, including $\boldsymbol{\beta}_{k'(-k)}$, is
{\small
\begin{align*}
p(\boldsymbol{\beta}_{k'k}\mid\boldsymbol{\beta}_{k'(-k)},\boldsymbol{X},\boldsymbol{Y}) & \propto p(\boldsymbol{\beta}_{k'k}\mid\boldsymbol{\beta}_{k'(-k)})
\prod_{i=1}^{N}\prod_{t=2}^{T_{i}}\left[\frac{e^{\zeta_{itk}-C_{itk}}}{1+e^{\zeta_{itk}-C_{itk}}}\right]^{Y_{itk}}\left[\frac{1}{1+e^{\zeta_{itk}-C_{itk}}}\right]^{1-Y_{itk}}\\
=p(\boldsymbol{\beta}_{k'k}\mid\boldsymbol{\beta}_{k'(-k)})&\prod_{i=1}^{N}\prod_{t=2}^{T_{i}}\left[e^{\kappa_{1itk}(\zeta_{itk}-C_{itk})}\int_{0}^{\infty}e^{-\omega_{1itk}(\zeta_{itk}-C_{itk})^{2}/2}p(\omega_{1itk})\,d\omega_{1itk}\right],
\end{align*}
}
where $\kappa_{1itk}=Y_{itk}-1/2$ and $\omega_{1itk}\sim\mathrm{PG}(1,0)$, for $k=1,\ldots,K$. Using this method, we can assume conditionally conjugate priors for the regression coefficients, $\boldsymbol{\beta}_{k'k} \sim N(m_0, V_0)$. Then, conditioning on the P\'olya-Gamma random variables $\omega_{1itk}$ and with some straightforward algebra, the posterior collapses into a single Gaussian kernel, leading to a two-step sampling scheme which involves a P\'olya-Gamma update for the latent variables $\omega_{1itk}$, followed by a joint Gaussian update for the regression coefficients $\boldsymbol{\beta}_{k'k}$. For additional details on this sampling step, see Appendix A of the Supplementary Material.

\subsubsection{Inference on the negative binomial probability parameters}
\label{section2.3.2}

Conditional on a latent state $\xi_{it} = k$, daily seizure counts are observed from a zero-inflated negative binomial (ZINB) emission distribution with dispersion parameter $r_k$, probability $\psi_{itk}$ and zero-inflation parameter $p_k$. In this section, we consider inference on the probability parameter $\psi_{itk}$, which is dependent on the covariates $\boldsymbol{X}_{it}$ through the logistic regression relationship of Eq~\eqref{psilogit}. We focus on inference for the regression coefficients $\boldsymbol{\rho}_k$ for the probability parameter and consider only the negative binomial mixture component of the ZINB distribution. 

Traditionally, Bayesian negative binomial regression is challenging due to the intractable form of the posterior and lack of a conjugate prior for the regression coefficients. \cite{pillowscott2012} describe an approach which circumvents these issues by taking advantage of P\'olya-Gamma auxiliary variables to derive a two-step closed-form Gibbs update for the regression coefficients. Due to the similar logistic link connecting the covariates to the parameter of interest, the sampling procedure for the negative binomial regression coefficients is similar to that of the transition probability regression coefficients. 

Conditional on $Z_{it} = 0$ and latent state $\xi_{it} = k$, the distribution of seizure counts for patient $i$ at time $t$ is negative binomial with
\begin{align*}
p(Y_{it}\mid & X_{it},\boldsymbol{r},\boldsymbol{\rho},\xi_{it}=k,Z_{it}=0)\propto(1-\psi_{itk})^{r_{k}}\psi_{itk}^{Y_{it}}.
\end{align*}
Then, leveraging the P\'olya-Gamma property of Eq~\eqref{pgkey}, we obtain the full conditional
{\small
\begin{align*}
p(\boldsymbol{\rho}_{k}\mid Y_{it},X_{it},\boldsymbol{r},\xi_{it}=k,Z_{it}=0) & =p(\boldsymbol{\rho}_{k})\,\prod_{i=1}^{N}\prod_{t=1}^{T_{i}}\frac{\left\{ exp(\boldsymbol{X}_{it}^{T}\boldsymbol{\rho}_{k})\right\} ^{Y_{it}}}{\left\{ 1+exp(\boldsymbol{X}_{it}^{T}\boldsymbol{\rho}_{k})\right\} ^{r_{k}+Y_{it}}}\\
 & \propto p(\boldsymbol{\rho}_{k})\,\prod_{i=1}^{N}\prod_{t=1}^{T_{i}}e^{\kappa_{2itk}\eta_{itk}}\int_{0}^{\infty}e^{-\omega_{2itk}\eta_{itk}^{2}/2}p(\omega_{2itk})\,d\omega_{2itk},
\end{align*}
}
where $\kappa_{2itk}=\frac{(Y_{it}-r_{k})}{2\omega_{2itk}}$, $\eta_{itk}=\boldsymbol{X}_{it}^{T}\boldsymbol{\rho}_{k}$ for brevity, and $\omega_{2itk}\sim\mathrm{PG}(Y_{it}+r_{k},0)$. The final expression combines to form a Gaussian kernel which allows us to derive a two-step Gibbs sampler for the negative binomial regression coefficients. Details of this sampler are provided in Appendix A of the Supplementary Material.

\subsubsection{Inference on the dispersion parameters}
For the update of the negative binomial dispersion parameters $r_k$, we adopt the data augmentation approach of \cite{zhou2015}, which exploits the relationship between the negative binomial distribution and the compound Poisson distribution. Conditional on $Z_{it} = 0$, the negative binomial counts can be rewritten as 
$Y_{it} =\sum_{l=1}^{L_{itk}}u_{itkl}$, with $u_{itkl}  \overset{iid}{\sim}\mathrm{Logarithmic}(\psi_{itk})$ and 
$L_{itk} \sim\mathrm{Poisson}(-r_{k}\,\mathrm{ln}(1-\psi_{itk}))$,
for $i=1,\ldots,N$ and $t=1,\ldots,T_i$ \citep{quenouille1949}. Assuming the prior $r_{k}\sim \mathrm{Gamma}(e,f)$, a two-step update for the dispersion parameter $r_k$ at each state $k = 1,\ldots, K$ iteratively draws
\begin{align*}
\left[L_{itk}\mid r_k \right] & \sim \mathrm{CRT}(Y_{it}I_{\xi_{it}}(k),r_{k}) \\ 
\left[r_{k}\mid L_{itk}, \psi_{itk}\right] & \sim \mathrm{Gamma}(e+\sum_{i=1}^{N}\sum_{t=1}^{T_{i}}L_{itk},f - \sum_{i,t} log(1-\psi_{itk})),
\end{align*}
where $\mathrm{CRT}$ is the Chinese restaurant table distribution.

\subsection{Posterior Inference}

For posterior inference, we design a MCMC algorithm that iteratively samples from the joint posterior distribution of all parameters, $\Theta= \left\{ \boldsymbol{\beta}_{k'},\boldsymbol{\gamma}, \boldsymbol{\pi},\xi_{it},\boldsymbol{\rho}_{k}, \boldsymbol{\delta},r_{k},p_{k}, Z_{it}\right\}$. Our strategy combines stochastic search methods for variable selection and the data augmentation techniques previously discussed. Our complete sampler is a Metropolis-within-Gibbs and is briefly summarized in Algorithm~\ref{alg:mcmc}. 

\begin{algorithm}
\label{alg:mcmc}
\SetAlgoLined
\SetKwInOut{Input}{Input}
\SetKwInOut{Output}{Output}
\Input{Iterations $S$, states $K$, data $\boldsymbol{X}$, $\boldsymbol{Y}$}
\Output{Posterior samples of model parameters $\Theta$}
\For{$s=1$ \KwTo $S$}{
    \For{$k=1$ \KwTo $K$}{
        Joint update of $({\boldsymbol{\beta}}, \boldsymbol{\gamma})$ via the stochastic search MCMC of  \cite{savitsky2011}, combined with the P\'olya-Gamma data augmentation scheme of \cite{polson2013}. This sampler performs an Add/Delete step on one component of $({\boldsymbol{\beta}}, \boldsymbol{\gamma})$ or a Swap step of two components, with equal probability. \\
        Joint update of $({\boldsymbol{\rho}}, {\boldsymbol{\delta}})$, similarly to the update of $({\boldsymbol{\beta}}, \boldsymbol{\gamma})$. \\
        Update $\boldsymbol{r}$ via the data augmentation method of \cite{zhou2015}.\\
        Update $\boldsymbol{p}$ from the full conditionals. \\
        Update $\boldsymbol{\pi}$ by sampling from the full conditionals. \\
    }
Update $\boldsymbol{\xi}$ via Scaled Forward-Backward algorithm \citep{scott2002}. \\
    \For{$i=1$ \KwTo $N$}{
        \For{$t=1$ \KwTo $T_i$}{
            Update $\boldsymbol{Z}$ from the full conditionals.\\
        }
    }
}
\caption{Sampler for ZINB-NHMM-BVS}
\end{algorithm}

Full details of the updates are given in Appendix A of the Supplementary Material. In our applications, starting values for all model parameters were initialized randomly. We discarded a fixed initial number of posterior samples, called the burn-in period, to control for the effect of initialization.  Posterior estimates of latent states for each patient were obtained via the posterior mode. The other model parameters were estimated via posterior means and 95\% credible intervals. Covariate effects were considered statistically significant if their marginal posterior probability of inclusion (MPPI) was greater than $0.5$ \citep{barbieri2004}. 

Finally, one common issue with MCMC methods for state-space models is label switching, which arises due to the invariance of the likelihood to permutations of the labels of the latent states \citep{scott2002}. To address this, at each iteration of the MCMC algorithm we calculate the negative binomial means as $\mu_{itk}= \frac{\psi_{itk}\,r_{k}}{1-\psi_{itk}}$ for $k=1,\ldots,K$, and enforce an ordering among the averaged state means as $\sum_{i=1}^{N}\sum_{t=1}^{T_{i}}\mu_{it1}<\sum_{i,t}\mu_{it2}<\ldots<\sum_{i,t}\mu_{itK}$.

\section{Simulation Study}
\label{Simulation Study}
We performed an extensive simulation study on the performance, accuracy, and sensitivity of our proposed model, ZINB-NHMM-BVS, on simulated seizure count data. We also compare the performance of our method for latent state classification and variable selection to the adaptive simulated annealing expectation maximization (ASA-EM) algorithm of \cite{hubin2019}, which performs simultaneous variable selection and inference on hidden states.

\subsection{Data generation}
\label{datagen}

We generated data by simulating $N = 100$ patients such that each patient had between 100 and 110 time points of data. If patients record daily seizure frequencies, this corresponds to about three months of data, which is a typical time between outpatient epilepsy visits. We set the number of hidden states to $K=3$ and set the corresponding dispersion parameters to $\boldsymbol{r} = (3,8,15)'$, the zero-inflation parameters to $\boldsymbol{p} = (0.7, 0.05, 0.01)'$ and the initial state distribution to $\boldsymbol{\pi} = (0.9, 0.08, 0.02)'$. We allowed covariates to affect the negative binomial parameters via construction \eqref{psilogit}. For each patient, $p=7$ covariates were designed to include a mix of both discrete and continuous features. Negative binomial regression coefficients were set as 
$\boldsymbol{\rho}_1 = (-0.7, -0.8, -0.8, 0, -0.8, -0.7, -0.7),$
$\boldsymbol{\rho}_2 = (-0.4, 0, 0, -0.4, 0, -0.7, -0.6)$ and $\boldsymbol{\rho}_3 = (0, -0.5, 0, -0.5, 0.5, 0.4, 0)$.
This approximately corresponds to an average of 0.2, 2.8, and 12.6 seizures per time unit at states 1, 2, and 3, respectively. We also allow all covariates to affect transitions from state 1 to 1, covariates \{$X_1$, $X_2$, $X_3$\} to affect transitions from state 1 to 2, covariates \{$X_2$, $X_3$, $X_7$\} to affect transitions from state 2 to 1, covariates \{$X_3$, $X_7$\} to affect transitions from state 2 to 2, and covariates \{$X_4$, $X_7$\} to affect both transitions from state 3 to 1 and from state 3 to 2. Non-zero transition regression coefficients were set as
$\left\{ \beta_{1,11},\beta_{2,11},\beta_{3,11},\beta_{4,11},\beta_{5,11},\beta_{6,11},\beta_{7,11}\right\} =3.5$, $\left\{ \beta_{1,12},\beta_{2,12},\beta_{3,12}\right\} =2.9$, $\left\{ \beta_{2,21},\beta_{3,21},\beta_{7,21}\right\} =2.4$,
$\left\{ \beta_{3,22},\beta_{7,22}\right\} =3.0, \left\{ \beta_{4,31},\beta_{7,31}\right\} =-2.9$ and $\left\{ \beta_{4,32},\beta_{7,32}\right\} =-2.5$.
Finally, we simulated the hidden states $\boldsymbol{\xi}$ and then generated the seizure counts $\boldsymbol{Y}$ from $\left[Y_{it}\mid \xi_{it} = k, \mathbf{r}, \mathbf{p}\right] \sim \mathrm{ZINB}(r_k, \psi_{itk},p_k)$. 

\subsection{Parameter settings}
\label{hyperparam}

When running MCMC chains, we set the following hyperparameters: We set non-informative Beta($1,1$) priors on the zero-inflation parameters, vague Gamma($.01,.01$) priors on the NB dispersion parameters and a non-informative Dir($1, \ldots, 1$) prior on the initial hidden state probabilities. We imposed Beta($1,5$) priors on the variable inclusion indicators for the transition probability regression and negative binomial regression coefficients, to allow for some sparsity in the selection model. We set mildly informative priors on the transition probability regression coefficients, $\boldsymbol{\beta}_{k'k} \sim N(\boldsymbol{0}_p,\mathrm{diag}(1))$, and the negative binomial regression coefficients, $\boldsymbol{\rho}_{k}\sim N(\boldsymbol{0}_p,\mathrm{diag}(1))$, to prevent the logistic terms \eqref{multilogit} and \eqref{psilogit} from taking on too extreme values (near 0 or 1). MCMC chains were run for 20,000 iterations, discarding the first 10,000 as burn-in, leaving 10,000 for inference. 

\subsection{Results}
We evaluated classification accuracy of the latent risk states using the following multi-class metrics:
\begin{align*}
\mathrm{Accuracy}{}=  \left(\sum_{k=1}^{K}\frac{\mathrm{TP}_{k}+\mathrm{TN}_{k}}{\mathrm{TP}_{k}+\mathrm{FP}_{k}+\mathrm{TN}_{k}+\mathrm{FN}_{k}}\right)/K\\
\mathrm{Precision}=  \left(\sum_{k=1}^{K}\frac{\mathrm{TP}_{k}}{\mathrm{TP}_{k}+\mathrm{FP}_{k}}\right)/K\\
\mathrm{Sensitivity}=  \left(\sum_{k=1}^{K}\frac{\mathrm{TP}_{k}}{\mathrm{TP}_{k}+\mathrm{FN}_{k}}\right)/K\\
\mathrm{Specificity}=  \left(\sum_{k=1}^{K}\frac{\mathrm{TN}_{k}}{\mathrm{TN}_{k}+\mathrm{FP}_{k}}\right)/K\\
\mathrm{F}_{1}\,\mathrm{score}=  2\cdot\frac{\mathrm{Precision}\,\mathrm{Sensitivity}}{\mathrm{Precision}+\mathrm{Sensitivity}},
\end{align*}
that weigh all latent states equally, and where TP$_k$, FP$_k$, TN$_k$, and FN$_k$ denote true positive, false positive, true negative, and false negative counts for state $k$. Using these macro-level metrics allows us to evaluate model performance in the presence of possible class (i.e.\ latent state) imbalance. 

Results on the inference of the latent states, averaged across 20 replicate datasets, are reported in the first row of Table \ref{varsellatentmetrics}, as the ``default'' model. Our method achieves good recovery of latent states with less than 5\% false positive and false negative rates. It also estimates other model parameters, such as dispersion, zero-inflation, and initial hidden states, with high accuracy and low mean-squared errors, see Table S3. Furthermore, it exhibits excellent performance on the selection of the covariates and the estimation of the corresponding regression coefficients. Table~\ref{varselsensitivity} (``default'' row) reports covariate selection results in terms of classification metrics, calculated based on a 0.5 threshold on the PPIs and averaged across 20 replicate datasets. Averaged estimates and root mean square errors (RMSE) on the corresponding regression coefficients are given in Tables S4 and S5. 

Additionally, we briefly investigated scalability to larger $p$ by augmenting the data matrix with 43 additional noise variables, for a total of $p=50$ covariates. Convergence was reached at 30,000 MCMC iterations and the resulting chain took approximately 8.7 hours to run. The model achieved high ($>$0.90) F$_1$ scores for classification of latent states as well as covariate selection, indicating that the method can scale up somehow to larger values of $p$.

\begin{table}
\centering
\caption{{\bf Simulation study (negative binomial): Sensitivity analysis. Classification metrics for latent state estimation.} Different slab variance parameters, $\Sigma_{\rho}$ and $\Sigma_{\beta}$, and different Beta priors on the  inclusion indicators are considered. The default set is $\Sigma_{\beta} = \mathrm{diag}(1)$, $g_{\beta} =1,h_{\beta} =5$, $\Sigma_{\rho} = \mathrm{diag}(1)$, and $g_{\rho} =1,h_{\rho} =5$, as described in Section ``Parameter settings''. Results are averaged across 20 replicated datasets. 
}
{\scriptsize
\begin{tabular}{c||ccccc}
\hline
Model & Acc. & Prec. & Sens. & Spec. & F$_1$\\
\hline
Default &  0.99 & 0.98 & 0.96 & 0.97 & 0.97 \\
$\Sigma_{\beta} = \mathrm{diag}(0.1)$ & 0.99 & 0.97 & 0.95 & 0.97 & 0.96\\
$\Sigma_{\beta} = \mathrm{diag}(10)$ &  0.99 & 0.97 & 0.96 & 0.97 & 0.97\\
$\Sigma_{\beta} = \mathrm{diag}(100)$ & 0.99 & 0.97 & 0.96 & 0.97 & 0.96\\
$g_{\beta} =1,h_{\beta} =1$ & 0.99 & 0.98 & 0.96 & 0.97 & 0.97 \\
$g_{\beta} =1,h_{\beta} =2$ & 0.99 & 0.98 & 0.96 & 0.97 & 0.97\\
$g_{\beta} =1,h_{\beta} =20$& 0.99 & 0.97 & 0.96 & 0.97 & 0.97\\
$\Sigma_{\rho} = \mathrm{diag}(0.1)$ & 0.99 & 0.98 & 0.96 & 0.97 & 0.97\\
$\Sigma_{\rho} = \mathrm{diag}(10)$ &  0.99 & 0.98 & 0.96 & 0.97 & 0.97\\
$\Sigma_{\rho} = \mathrm{diag}(100)$ & 0.99 & 0.98 & 0.96 & 0.97 & 0.97\\
$g_{\rho} =1,h_{\rho} =1$ & 0.99 & 0.98 & 0.96 & 0.97 & 0.97\\
$g_{\rho} =1,h_{\rho} =2$ & 0.99 & 0.98 & 0.96 & 0.97 & 0.97\\
$g_{\rho} =1,h_{\rho} =20$ & 0.99 & 0.98 & 0.96 & 0.97 & 0.97\\
\hline
\end{tabular}
}
\label{varsellatentmetrics}
\end{table}

\begin{table}
\caption{{\bf Simulation study (negative binomial):  Sensitivity analysis. Classification metrics for covariate selection.} Different slab variance parameters of the regression coefficients, $\Sigma_{\rho}$ and $\Sigma_{\beta}$, and different Beta priors on the variable inclusion indicators are considered. The default set is $\Sigma_{\beta} = \mathrm{diag}(1)$, $g_{\beta} =1,h_{\beta} =5$, $\Sigma_{\rho} = \mathrm{diag}(1)$, and $g_{\rho} =1,h_{\rho} =5$, as described in Section ``Parameter settings''. ``\% included'' refers to the prior expected number of included variables, and ``\# Selected'' refers to the number of variables selected by the median probability model. Results are averaged over 20 simulated datasets. The true data-generating model has 19 true non-zero covariates affecting Markov transitions, and 14 non-zero covariates affecting emissions. }
\centering
{\scriptsize
\vskip 1mm
{\it {Transitions}}
\vskip 1mm
\begin{tabular}{c||cccccccc}
\hline
& \% included & \# Selected & FNR & FPR & Prec. & Sens. & Spec. & F$_1$ \\
\hline
Default & 17\% & 20.60 & 0.02 & 0.09 & 0.91 & 0.98 & 0.91 & 0.94  \\
$\Sigma_{\beta} = \mathrm{diag}(0.1)$ & 17\% & 26.50 & 0.05 & 0.37 & 0.69 & 0.95 & 0.63 & 0.80 \\
$\Sigma_{\beta} = \mathrm{diag}(10)$ & 17\% & 20.65 & 0.04 & 0.10 & 0.89 & 0.96 & 0.90 & 0.92 \\
$\Sigma_{\beta} = \mathrm{diag}(100)$ & 17\% & 24.55 & 0.02 & 0.26 & 0.79 & 0.98 & 0.74 & 0.86 \\
$g_{\beta} =1,h_{\beta} =1$ & 50\% & 41.90 & 0.00 & 1.00 & 0.45 & 1.00 & 0.00 & 0.62 \\
$g_{\beta} =1,h_{\beta} =2$ & 33\% & 26.60 & 0.01 & 0.34 & 0.73 & 0.99 & 0.66 & 0.83 \\
$g_{\beta} =1,h_{\beta} =20$ & 5\% & 19.10 & 0.05 & 0.04 & 0.95 & 0.95 & 0.96 & 0.95 \\
\hline
\end{tabular}

\vskip 1mm
{\it {Emissions}} 
\vskip1mm
\begin{tabular}{c||ccccccccc}
\hline
& \% included & \# Selected & FNR & FPR & Prec. & Sens. & Spec. & F$_1$ \\
\hline
Default & 17\% & 13.95 & 0.02 & 0.04 & 0.98 & 0.98 & 0.96 & 0.98 \\
$\Sigma_{\rho} = \mathrm{diag}(0.1)$ & 17\% & 14.00 & 0.02 & 0.04 & 0.98 & 0.98 & 0.96 & 0.98 \\
$\Sigma_{\rho} = \mathrm{diag}(10)$ & 17\% & 13.95 & 0.02 & 0.04 & 0.98 & 0.98 & 0.96 & 0.98 \\
$\Sigma_{\rho} = \mathrm{diag}(100)$ & 17\% & 14.05 & 0.02 & 0.05 & 0.98 & 0.98 & 0.95 & 0.98 \\
$g_{\rho} =1,h_{\rho} =1$ & 50\% & 16.60 & 0.00 & 0.38 & 0.85 & 1.00 & 0.62 & 0.91 \\
$g_{\rho} =1,h_{\rho} =2$ & 33\% & 15.05 & 0.01 & 0.16 & 0.93 & 0.99 & 0.84 & 0.96 \\
$g_{\rho} =1,h_{\rho} =20$ & 5\% & 13.50 & 0.04 & 0.00 & 1.00 & 0.96 & 1.00 & 0.98 \\
\hline
\end{tabular}
\label{varselsensitivity}
}
\end{table}

\subsection{Sensitivity analysis}
Due to the hierarchical structure of our HMM, there are several hyperparameters that need to be set by the user before running the MCMC. In Section ``Parameter settings'', we provide guidelines for specification of non-informative or vague priors on the zero-inflation parameters, the dispersion parameters and the initial hidden state probabilities. We performed sensitivity analysis of the beta priors on the variable inclusion indicators and the Gaussian slab components of the {\it spike-and-slab} priors \eqref{betaspikeandslab} and \eqref{rhospikeandslab} on the regression coefficients, $\bm\beta_{k'k}$ and $\bm\rho_k$. Results in terms of the classification metrics are reported in Tables \ref{varsellatentmetrics} and \ref{varselsensitivity}, for latent state estimation and covariate selection, respectively. As expected, we found that varying the prior probability of variable inclusion affects the number of covariates selected into the model, in both the transition and emission components. Based on F$_1$ score, a good balance between FPR and FNR can be achieved by setting the prior probability of variable inclusion to be low, centered around 5--30\%. Overall, classification of latent risk states as assessed by the macro-level metrics was robust across the different prior specifications (Table~\ref{varsellatentmetrics}). 

Furthermore, we found that varying the slab variances $\Sigma_{\rho}$ of the negative binomial regression parameters $\boldsymbol{\rho}$ had minimal effect on both the covariates selected as well as the classification of the latent risk states. This was not the case for the slab variances $\Sigma_{\beta}$ of the transition probability regression parameters $\boldsymbol{\beta}_{k'}$. Setting the diagonal elements of the covariance matrix $\Sigma_{\beta}$ to extreme values increases the false positive rate for variable selection. However, classification of latent states remained robust even when the false positive rate for variable selection was high. 

In the above simulations, variable selection performance was assessed based on the median probability model, i.e.\ covariates were selected if their $\mathrm{MPPI}>0.5$. Furthermore, we investigated alternative ways to variable selection. For example, the most probable model chooses the model with the largest probability mass. In our implementation, each individual transition and state is treated separately and the most probable model is chosen based on the model that is visited the most often. 
In contrast, model selection based on the information criteria AIC or BIC chooses the model based on the MCMC iteration that minimizes the information criterion.
We provide a comparison of these various approaches in Table~\ref{tab:varselcompare}. Based on these simulations, the median probability model has very similar performance compared to the most probable model, while the models chosen by AIC or BIC lag behind both the median and most probable models.

Finally, we investigated the impact of various effect sizes on model performance. Specifically, we multiplied the effect sizes specified in Section~\ref{datagen} by a scale factor $F \in \{0.2,0.4,0.6,0.8,1\}$ and reassessed variable selection and latent state classification performances. Results are presented in Table S6 and indicate a measure of robustness even with the smallest effect sizes.

\begin{table}
\caption{{\bf Simulation study (negative binomial): Sensitivity analysis. Classification metrics for covariate selection, using different model selection approaches.} Results are averaged across 20 replicates each. ``\# Selected'' refers to the total number of variables selected by each model selection approach.}
\footnotesize
\centering
{\it Transitions}
\vskip 1mm
\begin{tabular}{c||ccccccc}
\hline
Model selection method & \# Selected & FNR & FPR & Prec. & Sens. & Spec. & F$_1$ \\
\hline
Median probability & 20.90 & 0.03 & 0.10 & 0.89 & 0.97 & 0.90 & 0.93 \\
Most probable & 20.55 & 0.03 & 0.10 & 0.90 & 0.97 & 0.90 & 0.93 \\
{AIC} & {21.10} & {0.08} & {0.16} & {0.84} & {0.92} & {0.84} & {0.87} \\
{BIC} & {19.50} & {0.09} & {0.10} & {0.89} & {0.91} & {0.90} & {0.90} \\
\hline
\end{tabular}
\vskip 1mm
{\it Emissions} 
\vskip 1mm
\begin{tabular}{c||cccccccc}
\hline
Model selection method & \# Selected & FNR & FPR & Prec. & Sens. & Spec. & F$_1$ \\
\hline
Median probability & 13.95 & 0.02 & 0.04 & 0.98 & 0.98 & 0.96 & 0.98 \\
Most probable & 13.95 & 0.02 & 0.03 & 0.99 & 0.98 & 0.97 & 0.98 \\
{AIC} & {14.45} & {0.02} & {0.11} & {0.95} & {0.98} & {0.89} & {0.96} \\
{BIC} & {13.90} & {0.03} & {0.04} & {0.98} & {0.97} & {0.96} & {0.97} \\
\hline
\end{tabular}
\label{tab:varselcompare}
\end{table}

\subsection{Comparison to ASA-EM}
\label{sec:asaem}

We compare the performance of our method to the non-homogeneous HMM modeling approach of \cite{hubin2019} that employs the adaptive simulated annealing expectation maximization (ASA-EM) algorithm for estimation. Similar to our approach, the ASA-EM approach allows covariates to affect both emissions and transitions. However, ASA-EM operates under the restriction that regardless of latent state, firstly, the same covariates affect the transitions and, secondly, the same covariates affect the emission distributions. The expectation maximization (EM) algorithm is used to infer unknown parameters, and the simulated annealing procedure is used for model exploration. The adaptive component of the method comes from the automatic learning of the model's tuning parameters over time. For additional details see \cite{hubin2019}.

The ASA-EM approach requires emission distributions to come from exponential families. Therefore, we re-simulated the count data using a Poisson emission distribution generated through a simple generalized linear model as
$Y_{it}\mid\xi_{it}=k, \mu_{itk}  \sim \mathrm{Poisson}(\mu_{itk})$, with
$\mathrm{log}(\mu_{itk}) =\mathbf{X}_{it}^{T}\boldsymbol{\zeta}_{k}$.
Again, we simulated data for $N = 100$ patients such that each patient had between 100 and 110 time points of data. We set the number of hidden states to $K=2$ and the initial state distribution to $\boldsymbol{\pi} = (0.9, 0.1)'$. For each patient, $p=15$ covariates were simulated and included a mix of both discrete and continuous features. The Poisson regression coefficients were fixed for each state at $\boldsymbol{\zeta}_1 = (-0.7, -0.7, 0, 0, -4, 0, -0.7, 0, 0, 0, 0, 0, 0, 0, 0)$ and
$\boldsymbol{\zeta}_2 = (0.5, -0.4, 0, 0, 0.7, 0, 0.5, 0, 0, 0, 0, 0, 0, 0, 0)$.
Thus, regardless of state, covariates \{$X_1$, $X_2$, $X_5$, $X_7$\} affected the emission distributions. This construction approximately corresponds to an average of 0.1 and 3.3 daily seizures at states 1 and 2, respectively. Finally, regardless of state we allowed covariates \{$X_2$, $X_3$, $X_5$, $X_7$\} to affect the transitions. 

Results reported here were obtained using the same hyperparameter settings described in the previous simulations. For the proposed ZINB-NHMM-BVS model, we considered three different choices for the variable inclusion hyperparameters in conjunction with the median probability model, to explore the effect of the beta prior. We also considered two alternative approaches to model selection, AIC and BIC. ASA-EM was run on 24 parallel threads for 3 epochs each and with parameters suggested in \cite{hubin2019}. AIC and BIC were used as the model selection criteria for ASA-EM, with a separate set of results reported for each. The choice of model selection criterion had a non-trivial impact on performance, and better performance gains may be realized if an information criterion more suitable for complex hierarchical models is used instead. Final results were averaged across 30 simulated datasets. 

\begin{table}
\centering
\caption{{\bf Simulation study (Poisson): Classification metrics for covariate selection.} The proposed model and the ASA-EM method are compared in terms of false positive and false negative rates, precision (Prec.), sensitivity (Sens.), specificity (Spec.), and F$_1$ score (F$_1$), for both the transition and emission components of the HMM. For the proposed ZINB model, different prior inclusion hyperparameters with the median model and alternative model selection approaches (AIC, BIC) were considered. Additionally, ``\% included'' refers to the prior expected number of included variables, and ``\# Selected'' refers to the number of variables selected by the models. Results are averaged across 30 replicates.
}
{\scriptsize
\vskip 1mm
{\it Transitions}
\vskip 1mm
\begin{tabular}{c||cccccccc}
\hline
Model & \% included & \# Selected & FNR & FPR & Prec. & Sens. & Spec. & F$_1$ \\
\hline
$g_{\beta} =1,h_{\beta} =3$ & 25\% & 8.87 & 0.00 & 0.04 & 0.91 & 1.00 & 0.96 & 0.95 \\
$g_{\beta} =1,h_{\beta} =5$ & 17\% & 8.20 & 0.00 & 0.01 & 0.97 & 1.00 & 0.99 & 0.98 \\
$g_{\beta} =1,h_{\beta} =10$ & 9\% & 8.30 & 0.00 & 0.01 & 0.97 & 1.00 & 0.99 & 0.98 \\
\hline
{ZINB, AIC} & {17\%} & {10.33} & {0.00} & {0.11} & {0.79} & {1.00} & {0.89} & {0.87} \\
{ZINB, BIC} & {17\%} & {9.47} & {0.00} & {0.07} & {0.86} & {1.00} & {0.93} & {0.92} \\
\hline
ASA-EM, AIC & - & 15.13 & 0.29 & 0.43 & 0.38 & 0.71 & 0.57 & 0.52 \\
ASA-EM, BIC & - & 8.80 & 0.52 & 0.22 & 0.48 & 0.48 & 0.78 & 0.57 \\
\hline
\end{tabular}
\vskip 1mm
{\it Emissions}
\vskip 1mm
\begin{tabular}{c||cccccccc}
\hline
Model & \% included & \# Selected & FNR & FPR & Prec. & Sens. & Spec. & F$_1$ \\
\hline
$g_{\rho} =1,h_{\rho} =3$ & 25\% & 11.00 & 0.00 & 0.14 & 0.75 & 1.00 & 0.86 & 0.85 \\
$g_{\rho} =1,h_{\rho} =5$ & 17\% & 9.83 & 0.00 & 0.08 & 0.83 & 1.00 & 0.92 & 0.90 \\
$g_{\rho} =1,h_{\rho} =10$ & 9\% & 9.57 & 0.00 & 0.07 & 0.85 & 1.00 & 0.93 & 0.92 \\
\hline
{ZINB, AIC} & {17\%} & {11.37} & {0.00} & {0.15} & {0.72} & {1.00} & {0.85} & {0.83} \\
{ZINB, BIC} & {17\%} & {10.57} & {0.00} & {0.12} & {0.78} & {1.00} & {0.88} & {0.87} \\
\hline
ASA-EM, AIC & - & 15.13 & 0.07 & 0.35 & 0.54 & 0.93 & 0.65 & 0.67 \\
ASA-EM, BIC & - & 14.00 & 0.05 & 0.29 & 0.62 & 0.95 & 0.71 & 0.73 \\
\hline
\end{tabular}
\label{freqcomparepoisson}
}
\end{table}

Variable selection performance metrics, including number of covariates selected, selection false negative rates, false positive rates and other metrics are reported in Table \ref{freqcomparepoisson}. Latent state classification performance metrics are reported in Table \ref{poissonlatentmetrics}. Overall, our proposed model achieves superior performance over ASA-EM in selecting covariates affecting the emission distributions, in selecting covariates that influence transitions, and in terms of overall classification of the latent states. The difference in performances could be attributed to several design choices which differ from ASA-EM. For instance, ZINB-NHMM-BVS implements a stochastic search algorithm based on an add-delete-swap sampler and modern data augmentation tools such as P\'olya-Gamma augmentation, which proved to be very effective in recovering regression coefficients affecting both the Markov transitions as well as emissions.

\begin{table}
\centering
\caption{{\bf Simulation study (Poisson): Classification metrics for latent states.} We compare (1) our proposed ZINB model with varying prior inclusion hyperparameters used in conjunction with the median model, and two alternative model selection approaches (AIC, BIC) and (2) ASA-EM, averaged across 30 simulated datasets. Metrics included are accuracy (Acc.), precision (Prec.), sensitivity (Sens.), specificity (Spec.) and F$_1$ score (F$_1$), with the first state set as the baseline state.}
\begin{tabular}{c||ccccc}
\hline
Model & Acc. & Prec. & Sens. & Spec. & F$_1$ \\
\hline
$g_{\beta} =1,h_{\beta} =3$ & 0.97 & 0.98 & 0.98& 0.95& 0.98\\
$g_{\beta} =1,h_{\beta} =5$ & 0.97 & 0.98& 0.98& 0.95& 0.98\\
$g_{\beta} =1,h_{\beta} =10$ & 0.97 & 0.98& 0.98& 0.95 & 0.98\\
\hline
$g_{\rho} =1,h_{\rho} =3$ & 0.97 & 0.98& 0.98& 0.95& 0.98\\
$g_{\rho} =1,h_{\rho} =5$ & 0.97 & 0.98& 0.98& 0.95 & 0.98\\
$g_{\rho} =1,h_{\rho} =10$ & 0.97 & 0.98& 0.98& 0.95 & 0.98\\
\hline
{ZINB, AIC} & {0.97} & {0.98} & {0.97} & {0.97} & {0.98} \\
{ZINB, BIC} & {0.97} & {0.99} & {0.97} & {0.97} & {0.98} \\
\hline
ASA-EM, AIC & 0.90 & 0.88 & 0.98 & 0.72 & 0.93 \\
ASA-EM, BIC & 0.89 & 0.87 & 0.98 & 0.69 & 0.92 \\
\hline
\end{tabular}
\label{poissonlatentmetrics}
\end{table}

\section{Dynamics of risk cycling in Dravet syndrome}
\label{Real-Data Application}
\subsection{Study subjects}
The Seizure Tracker$^{TM}$ system contains patient-reported data on seizure timestamps, seizure characteristics, triggers, and medications recorded by more than $30,000$ people with epilepsy across the world. Deidentified and unlinked data on all patients with self-identified Dravet syndrome in the SeizureTracker database between 2007 -- 2020 were exported from SeizureTracker.com on February 26, 2020. At the time of export, a total of 133 people with Dravet syndrome, with ages ranging between 2 months and 47 years, recorded $34,431$ GTCs between December 1, 2007 and February 26, 2020, spanning $141,499$ patient-days of data with an average of $1,064$ days and $259$ GTCs recorded per patient. Demographic information is reported in Table~\ref{patdemographics}.

\begin{table}
\centering
\caption{{\bf Characteristics of Dravet syndrome study sample ($n=133$).}
For continuous variables, mean (SD) are shown. For categorical variables, proportion of seizures ($^{\dag}$) or patients ($^{\ddag}$) is shown. $^*$Only generalized tonic-clonic seizures were considered. $^{\S}$Status epilepticus defined as seizure with duration $>$5 minutes.}
\scriptsize
\begin{tabular}{|l|r|}
\hline
Characteristic & Value \\
\hline
Demographic characteristics & \\
\hline
\hspace{12pt} Age in years, mean (SD) & 8.25 (6.55) \\
\hspace{12pt} Male sex$^{\ddag}$ & 51.1\% \\
\hline
Seizure triggers$^{\dag}$ & \\
\hline
\hspace{12pt} Change in medications & 7.1\% \\
\hspace{12pt} Tiredness & 13.3\% \\
\hspace{12pt} Overheating & 3.7\% \\
\hspace{12pt} Illness & 5.8\% \\
\hspace{12pt} Bad mood & 2.9\% \\
\hline
Seizure burden$^*$ & \\
\hline
\hspace{12pt} Duration in minutes, mean (SD) & 1.94 (5.42)\\
\hspace{12pt} Status epilepticus$^{\S}$ & 4.8\% \\
\hline
Use of VNS therapy$^{\ddag}$ & 21.1\% \\
\hline
Seizure frequency$^*$ & \\
\hline
\hspace{12pt} Seizures per day, mean (SD) & 0.24 (0.92) \\
\hspace{12pt} Mean proportion of seizure-free days & 85.4\% \\
\hspace{12pt} Mean proportion of days with one seizure & 10.2\% \\
\hspace{12pt} Mean proportion of days with 2+ seizures & 4.4\% \\
\hline
Anti-seizure medication usage$^{\ddag}$  & \\
\hline
\hspace{12pt} Benzodiazepines & 43.6\% \\
\hspace{12pt} Valproic acid & 41.4\% \\
\hspace{12pt} Stiripentol & 24.8\% \\
\hspace{12pt} Levetiracetam or brivaracetam & 21.8\% \\
\hspace{12pt} Topiramate & 20.3\% \\
\hspace{12pt} Cannabidiol & 18.0\% \\
\hspace{12pt} Tetrahydrocannabinols & 7.5\% \\
\hspace{12pt} Zonisamide & 6.8\% \\
\hspace{12pt} Phenobarbital or phenobarbital-containing compounds & 6.0\% \\
\hspace{12pt} Felbamate & 4.5\% \\
\hspace{12pt} Ethosuximide or methsuximide & 3.8\% \\
\hspace{12pt} Lacosamide & 3.0\% \\
\hspace{12pt} Carbonic anhydrase inhibitors & 2.3\% \\
\hspace{12pt} Lamotrigine & 2.3\% \\
\hspace{12pt} Prednisolone & 1.5\% \\
\hspace{12pt} Triple bromide or potassium bromide & 1.5\% \\
\hspace{12pt} Carbamazepine, eslicarbazepine, oxcarbazepine & 1.5\% \\
\hspace{12pt} Phenytoin & 0.8\% \\
\hspace{12pt} Primidone & 0.8\% \\
\hspace{12pt} Perampanel & 0.8\% \\
\hline
Other medications$^{\ddag}$ & \\
\hline
\hspace{12pt} Verapamil & 3.0\% \\
\hspace{12pt} Risperidone & 2.3\% \\
\hspace{12pt} Vitamin B6 & 2.3\% \\
\hline
\end{tabular}
\label{patdemographics}
\end{table}

\subsection{Data pre-processing}
A total of 618 patients with Dravet syndrome were initially identified from 2007 -- 2020 from the Seizure Tracker database. Patients were excluded if their SeizureTracker.com account was used for less than 30 days, if fewer than 20 GTCs were recorded, or if patients averaged less than one seizure per year. This allowed for exclusion of patients who did not consistently record seizures and to increase the precision of estimates. Patients with missing or invalid dates of birth or sex were also excluded.  Invalid seizure durations were imputed as the mean duration of that seizure type for the patient or, if unavailable, the mean duration of that seizure type for all patients. Continuous covariates were normalized to the 0--1 range. 

Individual medication records were excluded if the medication name or recorded start date were missing or invalid. For medications with missing end dates, the end date was imputed as three weeks (21 days) after the medication start date. A three-week imputation was considered reasonable as ASMs are often uptitrated over 2--4 weeks to attain a therapeutic dosage \citep{panayiotopoulos2005chapter}. Rescue medications were not included in analysis. Generic and brand name medications were analyzed in the same class \citep{privitera2016generic,vossler2016aes}. 
 
The influence of $p=36$ clinical covariates on changes in seizure risk were evaluated, including 20 classes of daily dosed ASMs, three adjuvant medications, nine common seizure triggers, vagus nerve stimulation (VNS) therapy usage, and three clinical characteristics (age, sex, and menstrual cycling). A complete list of medication classes, seizure triggers, and clinical characteristics is reported in Table S7.

\subsection{Seizure risk states in Dravet syndrome}
We applied our model to GTCs in Dravet syndrome in order to evaluate the characteristics and dynamics of seizure risk cycles in patients with this condition, while simultaneously identifying ASMs and triggers associated with changes in GTC risk. In addition to the $p=36$ clinical covariates from Seizure Tracker, we also estimated the effect of the intercept in both the transition and emission components of the model. Results were obtained by running  MCMC chains for 20,000 iterations with 10,000 sweeps as burn-in. Hyperparameters were set to be weakly informative as described in the simulation studies. Convergence of the MCMC chains was assessed via visual inspection of trace plots and auto-correlation plots, and by the Geweke's diagnostic test \citep{geweke1992}. Convergence results are shown for a few selected parameters that are sampled via P\'olya-Gamma data augmentation in Appendix B of the Supplementary Material. The optimal number of states, $K$, was chosen based on minimization of the deviance information criterion \citep[DIC, defined as in][]{gelman2013} over a grid of possible values $K \in \{2,3,4,5,6,7\}$. DIC values were obtained using the BEST approach of selecting covariates based on the median probability model (as opposed to the FULL model which includes all covariates or the NULL model which includes no covariates). More specifically, let $\hat{\theta}_{\mathrm{Bayes}}=\mathrm{E}(\theta|y)$ be the posterior mean and $\theta^{s}$ be posterior draws at MCMC iteration $s=1,\ldots,S$. Then, the DIC is obtained as
$$\mathrm{DIC}=-2\mathrm{\,log}\,p(y|\hat{\theta}_{Bayes})+2p_{\mathrm{DIC}},$$
where 
$$p_{\mathrm{DIC}}=2\left(\mathrm{log}\,p(y|\hat{\theta}_{Bayes})-\frac{1}{S}\sum_{s=1}^{S}\mathrm{log}\,p(y|\theta^{s})\right).$$

\begin{figure}
\centering
\caption{{\bf Dravet syndrome Seizure Tracker data analysis: Distributions of daily GTC counts in each latent risk state.}
These are estimated in our model by first estimating the latent states $\boldsymbol{\xi}$ via the posterior mode and then plotting the empirical distribution of the observed daily counts in each state.}
\includegraphics[scale=0.5]{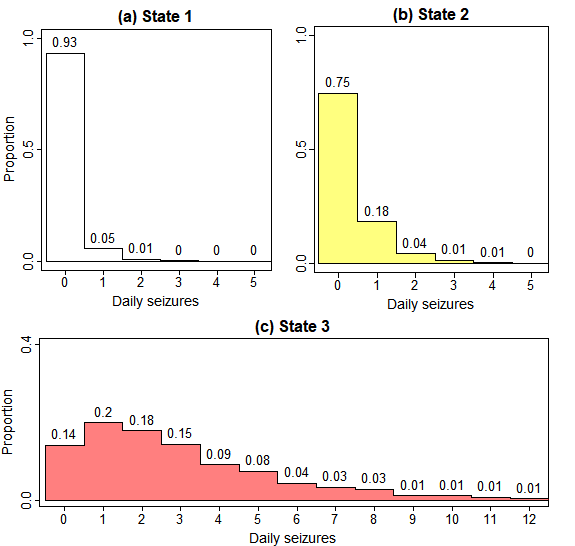}
\label{states}
\end{figure}

We found evidence for $K=3$ distinct seizure risk states in Dravet syndrome based on minimization of the DIC. These three states can be interpreted with respect to their pro-ictal tendency as states in which the patient is at low, moderate, or high risk for GTCs. The empirical distribution of seizures within each of the identified states provides information on the distribution of seizures that a patient might expect on days in each state (Fig~\ref{states}). State 1 (``low'' risk) was associated with the lowest seizure frequency. Patients in this state could expect to be seizure-free from GTCs on 93\% of days in this state, with a mean GTC seizure frequency of 1.31 (SD, 0.82) on days when at least one GTC occurred. Patients with Dravet syndrome spent the majority of time (63.8\% of all days) in this low risk state. In state 2 (``moderate'' risk), patients also had reasonable expectation for freedom from GTCs, and could expect to be seizure-free from GTCs on 75\% of days in this state, with a mean GTC seizure frequency of 1.42 (SD, 0.88) on days when at least one GTC occurred. On average, patients were in this moderate risk state about 34.3\% of all days. In state 3 (``high'' risk), patients had a dramatically higher risk for GTCs. In this state, patients could expect to have at least one GTC with more than 85\% likelihood, and at least two GTCs per day on two-thirds of days spent this state. These estimates may be informative for counseling patients on seizure-related injury and death. For example, given the association between GTCs and SUDEP risk, patients in risk state 3 would be predicted to be at extremely high risk of SUDEP on days in this state. Patients spent the least amount of time (1.9\% of all days) in this high risk state. 

\subsection{Mean sojourn time and volatility of risk states in Dravet syndrome}
Our inference yields information on the mean sojourn times (MST) for each seizure risk state, which provide a measure of the expected duration of time that a patient with Dravet syndrome is likely to spend in each seizure risk state before transitioning out of the state. This quantity is empirically estimated based on the posterior mode of the latent risk states (see Figure S3). The MST for the low risk state was 117.6 days (about 3.9 months) with an interquartile range (IQR) of 12 to 123 days, indicating large variance in the sojourn time for the low risk state. We found that the sojourn time of the low risk state was $<$12 months in 91.4\% of patients. The MST for the moderate risk state  was 59.7 days with an IQR of 3.0 to 53.5 days. In contrast, the MST for the high risk state was a shorter 5.8 days (IQR, 2 to 6 days). 

\begin{figure}
\centering
\caption{{\bf Dravet syndrome Seizure Tracker data analysis: Mean transition matrix.}
These are estimated in our model by first calculating $\mathrm{Pr}(\xi_{it}=k\mid\xi_{i,t-1}=k',\cdot)$ via the multinomial logistic equation \eqref{multilogit} with coefficients $\boldsymbol{\beta}_{k'}$ estimated by the posterior mean, then averaging across all patients and days for each transition. Standard deviations (SD) are shown in parentheses.}
\includegraphics[scale=0.5]{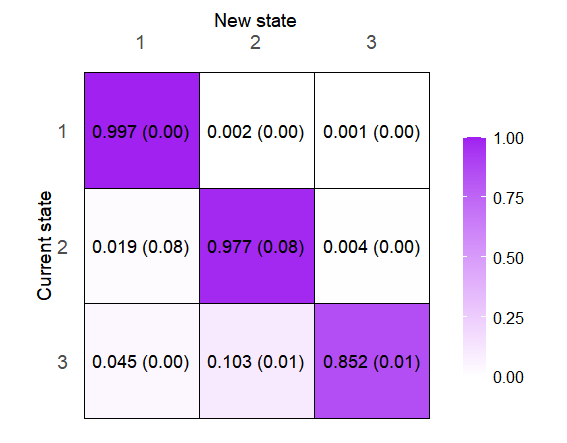}
\label{avgtp}
\end{figure}

We also investigated the volatility of the dynamics of transitions between seizure risk states, based on the distribution of the estimated transition probabilities \eqref{multilogit}, which provide estimates of the likelihood a patient will be in a certain state on the next day. We found that seizure risk states in Dravet syndrome have a strong tendency toward self-transitioning (i.e., from State X to State X) on a daily timescale. In other words, once a patient enters a given risk state, s/he is more likely to remain in that state the following day, than to transition to a different state. The self-transitioning property was strongest for state 1, less for state 2, and least for state 3; for example, see the average transition matrix in Fig~\ref{avgtp}. Figure S4 also shows the distribution of transition probabilities for each individual transition. 

\subsection{Incorporation of external clinical covariates improves accuracy of seizure risk estimation}
\begin{figure}
\centering
\caption{{\bf Dravet syndrome Seizure Tracker data analysis: Expected number of seizures for the sequence of daily seizure counts from one patient.} These are estimated with (a) the proposed non-homogeneous HMM (ZINB-NHMM-BVS), (b) a baseline homogeneous HMM (ZINB-HHMM), and (c) the method of \cite{chiang2018} (ZIP-NHMM). Values are shown on the log scale. Unlike ZIP-NHMM and ZINB-HHMM, which produce state-specific estimates of risk, ZINB-NHMM-BVS results in subject- and state-specific estimates, obtained as the estimated NB mean parameters, $\mu_{itk}= \frac{\psi_{itk}\,r_{k}}{1-\psi_{itk}}$, with overdispersion parameters $r_k$ estimated via posterior mean and probabilities $\psi_{itk}$ calculated via \eqref{psilogit}, with latent risk states $\xi_{it}$ estimated by posterior mode and regression coefficients $\boldsymbol{\rho}_k$ via posterior mean. The heatmap at the bottom of each panel shows the most likely sequence of risk states for that method.}
\includegraphics[scale = 0.35]{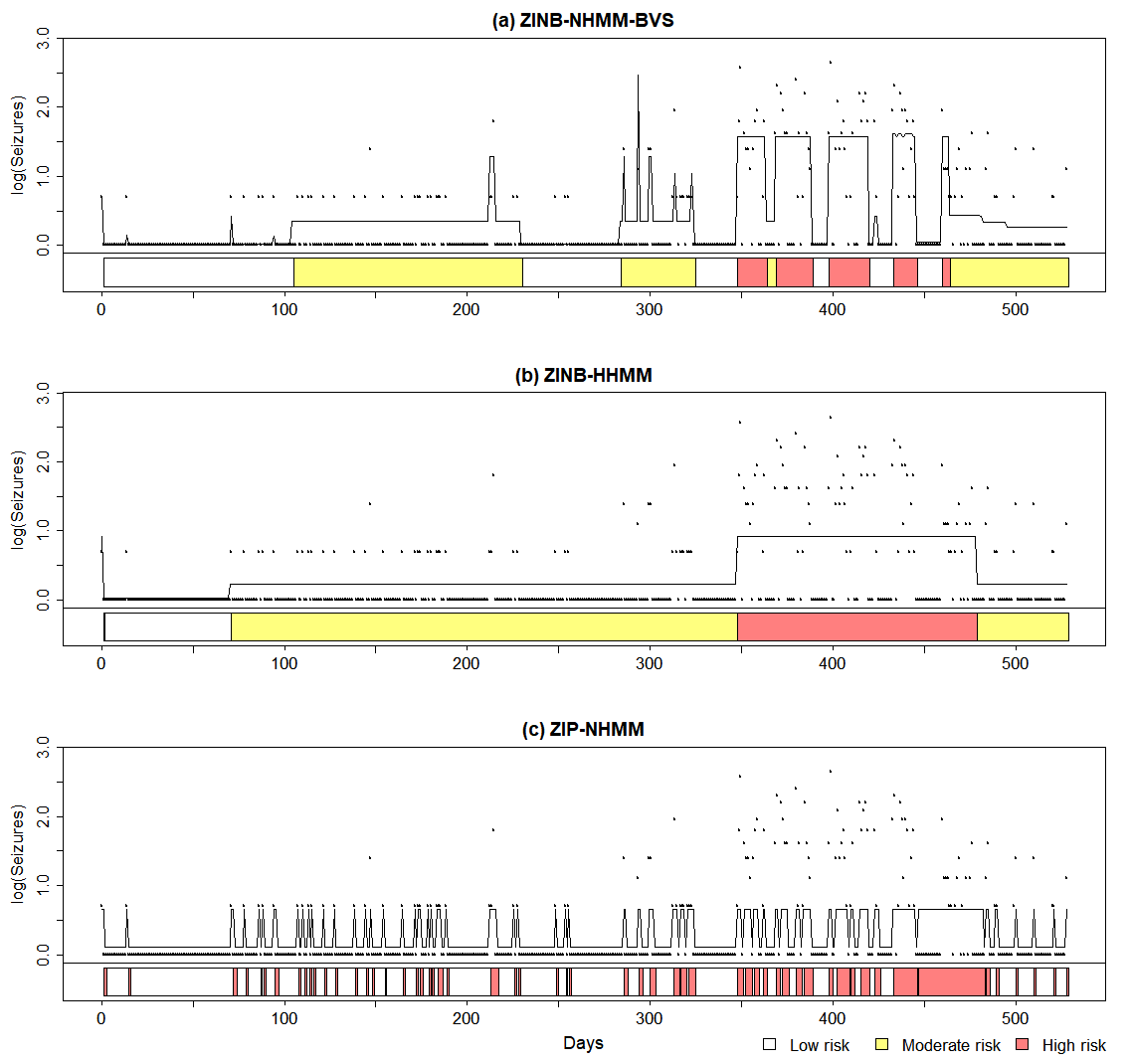}
\label{matchedmeans}
\end{figure}

Next, we show that accounting for external variables on modulatory factors improves accuracy of the estimates for seizure risk states. Fig~\ref{matchedmeans}a shows the expected number of GTCs based on our proposed model (ZINB-NHMM-BVS) for the sequence of daily seizure counts from a randomly selected patient with Dravet syndrome. The expected number of GTCs based on a {\it homogeneous} HMM (ZINB-HHMM), in which patient-level covariates are {\it not} incorporated into estimation of the mean seizure process $\mu_{it}$, is shown for comparison in Fig~\ref{matchedmeans}b. We found that the use of a non-homogeneous HMM corresponds to a 8.6\% improvement in mean absolute error, defined as the average absolute difference between the observed response $Y_{it}$ and the mean value estimated by the model $\hat{\mu}_{it}$, over a homogeneous HMM that does not incorporate covariates into the seizure risk estimate. In terms of deviance information criterion, the DIC of the proposed method (ZINB-NHMM-BVS) was 119,790, compared to the DIC of the homogeneous model (ZINB-HHMM) which was a higher 138,854. These metrics suggest that accounting for clinical variables into statistical algorithms for estimating seizure cycles may improve the accuracy of seizure risk estimation, particularly compared to methods which do not consider external variables into their estimates of seizure risk, and emphasizes the utility of incorporating multimodal data streams into seizure risk algorithms. 

We also illustrate the methodological advances of the current model compared to our original model in \cite{chiang2018}, which we refer to as ZIP-NHMM, and demonstrate different situations in which the two models may be used. First, as shown in Fig~\ref{matchedmeans}c, using ZIP-NHMM, the state-specific estimates of the expected number of seizures at each risk state ($\mu_k$) are fixed and do not vary based on patient-level covariates. In contrast, ZINB-NHMM-BVS produces subject- and state-specific estimates of the expected number of seizures in each risk state ($\mu_{itk}$, Eq~\ref{psilogit}). This will provide improved estimation of the expected number of seizures at each time point in situations where the distribution of seizure counts occurring during a specific risk state is more variable. The difference between the two models can be seen by contrasting Fig~\ref{matchedmeans}a and c. During the first 300 days, the estimated mean process is better for ZIP-NHMM, as the patient has a less variable range of seizure counts occurring in these risk states (here, 0 or 1 seizures per day) (Fig~\ref{matchedmeans}c). In the following 200 days, the patient has a more variable range of seizure counts that may occur, with better estimation of the mean process by ZINB-NHMM-BVS (Fig~\ref{matchedmeans}a). Second, the current model improves on the ability to distinguish between distinct risk states in datasets when the number of events is small. As shown in Fig~\ref{matchedmeans}a and c, ZINB-NHMM-BVS distinguishes between low, moderate, and high risk states, whereas ZIP-NHMM groups the moderate and high risk states into a single high risk state. This becomes relevant in data with increased temporal granularity, such as intracranial electrocorticography data or real-time wearable data \citep{chiang2021evidence}, for which counts in smaller time bins are often close to zero. 

We conduct additional experiments to compare the performance of ZINB-NHMM-BVS to the method of \cite{hubin2019}, ASA-EM, as well as to various submodels of ZINB-NHMM-BVS that utilize simpler variable selection schemes for the transition probability regression coefficients, on the Dravet syndrome Seizure Tracker dataset. These results are presented in Appendix C of the Supplementary Material.

\subsection{Drivers of risk cycles in Dravet syndrome}
Next, we demonstrate how the methodological addition of variable selection to our model results in a method that can be used to study drivers of seizure risk cycles. By allowing for covariate selection in our model, in particular, we can distinguish between clinical variables that may have differential effects on seizure risk, which may either have an acute effect on the number of seizures the patient is likely to have at time $t$, or a more subacute effect on how high risk the patient is likely to become at time $t+1$. The rationale for this model development is to capture the fact that some factors, such as electrolyte abnormalities or missing a medication dose, may exert a more acute influence on the expected number of seizures at the current time point; other factors, such as hormonal cycles, may exert a more long-term influence on the probability of being high risk at a future time point.

\begin{figure}
\centering
\caption{{\bf Dravet syndrome Seizure Tracker data analysis: Marginal posterior probabilities of inclusion (MPPI).} MPPI are shown for (a) transition probability regression coefficients $\boldsymbol{\beta}_{k'}$ by covariate, and for (b) negative binomial regression coefficients $\boldsymbol{\rho}_k$ by state and covariate. Variable inclusion criterion is set at MPPI $>0.5$ and is indicated by the red line. Covariates are split into three broad categories: patient demographics (gray area), seizure triggers (pink area), and medication classes (blue area). Left-most covariate (position 1) is the bias term. Baseline state is $k = 3$. A key indicating the variable name corresponding to each covariate index can be found in Table S7.} 
\includegraphics[scale = 0.4]{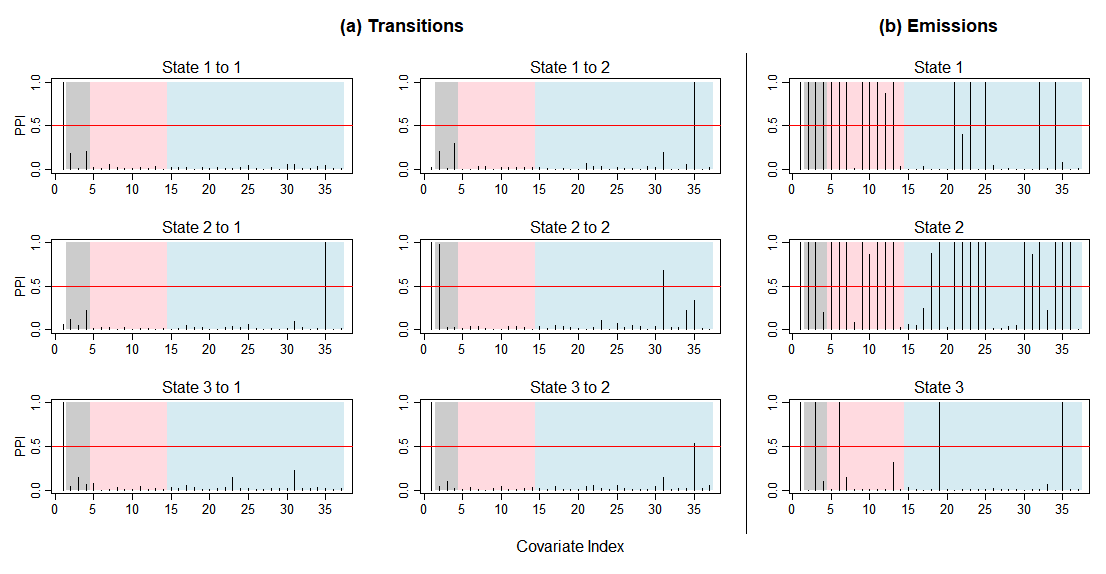}
\label{mppi}
\end{figure}

Marginal posterior probabilities of inclusion (MPPI) for transition probability regression coefficients $\boldsymbol{\beta}_{k'}$, by covariate, and for negative binomial regression coefficients $\boldsymbol{\rho}_k$, by state and covariate, are shown in Fig~\ref{mppi}. This figure illustrates the significance of covariate effects on either (a) transitions from one risk state to another or (b) the number of seizures given the current risk state. The reference state was set to $k = 3$ (the high seizure risk state) such that the regression coefficients corresponding to transitions into this state were zero, to maintain model identifiability. Selection of a different reference state resulted in a similar model fit based on DIC and variable inclusion agreement for the majority of regression coefficients. Selected covariates (i.e., $>50\%$ marginal posterior probability) for state transitions are reported in Table \ref{tpvals}, together with posterior means and credible intervals of the corresponding regression coefficients. For patients currently in low or moderate seizure risk states (1 or 2), the use of cannabidiol (CBD) was associated with a greater likelihood of remaining in those states, as opposed to worsening to the higher risk state 3. For patients currently in risk state 2 (moderate risk for GTCs), older patient age and treatment with zonisamide were associated with an increased likelihood of remaining at the same risk the following day, as opposed to worsening to the higher risk state 3. 

\begin{table}
\centering
\caption{{\bf Dravet syndrome Seizure Tracker data analysis: Posterior means and SD for transition probability regression coefficients, $\boldsymbol{\beta}_{k'k}$, of selected covariates.} These are reported with 95\% credible intervals (defined as the interval $\text{\ensuremath{\left[a,b\right]}}$ such that Pr($a \leq \beta_{j,k'k} \leq b \mid data$) = 0.95) and marginal posterior probabilities of inclusion (MPPI). Only covariates with MPPI $>0.5$ are shown. Baseline state is $k=3$. }
\begin{tabular}{cccccccc}
\hline
Transition & Covariate & Post.\ mean (SD) & MPPI & 95\% CI \\
\hline
1 $\rightarrow$ 1 & Intercept & 6.65 (0.16) & 1.00 & (6.38, 6.95)\\
2 $\rightarrow$ 2 & Intercept & 4.98 (0.19) & 1.00 & (4.64, 5.36)\\
3 $\rightarrow$ 1 & Intercept & -2.85 (0.33) & 1.00 & (-3.54, -2.23)\\
3 $\rightarrow$ 2 & Intercept & -2.04 (0.33) & 1.00 & (-2.89, -1.58)\\
2 $\rightarrow$ 2 & Age & 2.99 (0.59) & 0.98 & (1.81, 4.16)\\
2 $\rightarrow$ 2 & Zonisamide & 5.47 (0.94) & 0.68 & (3.29, 7.06)\\
1 $\rightarrow$ 2 & Cannabidiol & 3.15 (0.49) & 1.00 & (2.33, 4.12)\\
2 $\rightarrow$ 1 & Cannabidiol & 5.50 (0.85) & 1.00 & (3.54, 6.62)\\
3 $\rightarrow$ 2 & Cannabidiol & -2.36 (0.70) & 0.53 & (-3.90, -1.13)\\
\hline
\end{tabular}
\label{tpvals}
\end{table}

Various triggers and ASMs were found to have an acute effect on the number of GTCs a patient could expect on a given day, given the current risk state (see  Table S8 and Figure S5). The influence of patient-level covariates on modulating the number of seizures on a given day was greatest for patients currently in a ``moderate'' risk state. For patients in this risk state, bad mood, sudden changes in daily medication regimen (e.g.\ missed/late medications or changing medications), illness, and tiredness were associated with a greater number of expected seizures that day. This is consistent with research finding that sleep and anxiety are relevant in patient self-prediction of seizures \citep{haut2007seizure}. Among ASMs, we found that perampanel, triple or potassium bromide, and verapamil were associated with reducing the expected number of GTCs in state 2. Preliminary evidence of efficacy of these medications as adjunctive therapies has been identified in other studies as well \citep{yoshitomi2019efficacy}. 

\section{Conclusion}
\label{Discussion}
In this paper, we have developed a rigorous Bayesian non-homogeneous hidden Markov modeling approach for estimating seizure risk based on observed seizure counts and demonstrated its usage for investigating drivers of risk cycling in epilepsy. The approach accounts for the overdispersion exhibited in daily seizure count data due to seizure unpredictability and natural variability. It also incorporates exogenous clinical covariates into the estimation of model parameters, allowing for a more precise assessment of seizure risk. Unlike existing approaches, we relax the assumption that clinical covariates that drive risk are known {\it a priori}, but rather simultaneously infer important drivers through variable selection priors. Our approach allows improved granularity of seizure risk assessment as well as an integrated framework for identifying drivers of risk cycles.  

The utility of this model is demonstrated through our application analyzing self-recorded seizures by people with Dravet syndrome through SeizureTracker.com, a widely used seizure diary. We demonstrate the presence of three distinct states in Dravet syndrome at different risk for GTCs. We have characterized the volatility of these states, showing that patients with Dravet syndrome tend to self-transition (i.e., remain in the same state once in the state). We have shown that ASMs and triggers may drive fluctuations in seizure risk through differential effects on either (1) future state transitions, or (2) expected seizure frequency given the current risk state. This suggests that drivers of risk cycles can either exhibit a short-term effect, by reducing the number of expected seizures in the current time-frame, or a longer term effect, by modulating which risk state the patient is likely to transition to in the future. We have shown that incorporating the influence of external modulatory variables into statistical inference on seizure risk, such as medications and seizure triggers, improves the accuracy of risk estimation, compared to estimation based on seizure counting data alone.

The usefulness of our model for characterizing risk cycles in Dravet syndrome suggests potential for broader utility in the field of seizure risk modeling in epilepsy. Estimation of the MST of the high GTC risk state may potentially be relevant for guiding preventative counseling in risk for SUDEP and seizure-related injury. In particular, mean estimates and confidence intervals for the MST of high risk states are useful for counseling patients on how much time the average patient can expect to remain at high risk for GTCs once they enter a high risk state. For patients in high GTC risk states (and increased risk for SUDEP and seizure-related injury), an estimated MST of 5.8 days may suggest a possible approach of counseling patients in this state to take heightened seizure precautions for at least one week. 

There are several limitations to this study. The medication data analyzed here is self-reported, and therefore results about the efficacy of pharmacological treatments on reducing seizure risk in Dravet syndrome should be considered with this caveat. This study primarily serves to validate our model's usage for investigating pharmacological relationships and extension to randomized clinical trial data is needed for rigorous studies of medication efficacy. Generally speaking, clinical applications of our method can directly improve patient care through a better understanding of the natural history of risk cycles in epilepsy, allowing clinicians to act proactively with drug interventions based on the understanding of expected fluctuations in risk cycles. Furthermore, the model can be used to understand the pharmacological effectiveness of therapies in modulating risk cycles in epilepsy. Although we have demonstrated our model's utility on patient-reported clinical seizure count data, our model is applicable to seizure counting data from any source, including intracranial, subdural, or scalp electroencephalography. 

In our model parametrization, the number of parameters can grow quickly as the number of predictors or latent states increases, potentially resulting in overfitting. This is especially true for the transition model, as a separate set of regression coefficients is specified for each individual transition. 
In our approach, this issue is mitigated by the use of sparse priors for covariate selection.

Recent studies have suggested that individual gene expression may affect the resistance of seizures to ASMs \citep{naimo2019}. In this context, our model is well-equipped to evaluate the effect and significance of varying genomic profiles on drug-refractory epilepsy. Furthermore, in the field of seizure forecasting and prediction, in addition to the predicted next risk state, the predicted absolute sequence of states may also be of clinical interest. Evaluation of how well the model performs in prospective on-line predictions of seizure risk will be of interest in future work, to determine contribution of this model to other seizure forecasting systems. 

Our approach could be extended to infer the optimal number of states $K$ via a sampling scheme that can accommodate posterior distributions of varied dimensions, for example reversible jump MCMC, though  at  the  expense  of  an  increase in computational cost. Our proposed model offers deep insights into the state-dependence of risk factors contributing to seizure propensity at both the Markov transitions and count emissions. Such inferences may provide great clinical value, for example through helping clinicians make more informed decisions regarding treatment plans for patients by leveraging knowledge of which medications or treatments are most effective conditional on the current seizure risk state of the patient. On the other hand, in situations where the clinical interpretation is not essential, a simpler model such as one of the submodels suggested in Appendix C should be preferred. As another example, a nonparametric approach to estimation of the count-valued emission distribution or transition probabilities would simultaneously improve flexibility and reduce the complexity of the model, though potentially at a cost of computation time and interpretability. For example, the general case of the method of \cite{canale2011} models the count observations via a kernel mixture of counts but, unlike our model, does not assess the effect of additional covariates on the response. A multivariate extension allows for joint modeling of count observations with continuous and categorical predictors via a multivariate rounded mixture of Gaussians. However, this introduces additional complexities into the model and further complicates interpretation.

\section*{Acknowledgments}
We thank the people living with Dravet syndrome using SeizureTracker.com, who allowed their de-identified data to be used in this analysis. 
Data utilized in the seizure count case study was provided by Seizure Tracker™ - \url{https://seizuretracker.com/}, and should be directly requested from Seizure Tracker, LLC. 
ETW is supported by a fellowship from the Gulf Coast Consortia, on the NLM Training Program in Biomedical Informatics and Data Science T15LM007093. SC is supported by the National Institute of Neurological Disorders and Stroke, National Institutes of Health (5R25NS070680-12). The contents are solely the responsibility of the authors and do not necessarily represent the views of the NIH.

\section*{Supplementary Material}

The software implementing our model, with setup instructions and codes replicating the simulation studies, can be found as online supplement. To request the decryption password from the authors please complete the form at \url{https://forms.gle/hc8UHsB3bSdisQX27}. The most updated version of the software can be dowloaded from GitHub at  \url{https://github.com/seizurerisk}. 


Appendix A of the online supplement contains a detailed description of the MCMC algorithm implemented for our proposed method. 

Appendix B contains example MCMC trace plots, auto-correlation plots, and results of Geweke's diagnostic tests to demonstrate convergence of the MCMC chain in the Seizure Tracker study, accompanied by a short discussion. 

Appendix C contains a comparison of our proposed method, ZINB-NHMM-BVS, to the approach of \cite{hubin2019}, ASA-EM, as well as to various submodels of ZINB-NHMM-BVS which employ simpler variable selection schemes for the transition probability regression coefficients on the Seizure Tracker dataset. 

Appendix D contains supplemental Tables and Figures.

\bibliographystyle{imsart-nameyear}
\bibliography{bib.bib}

\end{document}